# QoE Assessment for SVC Live Distribution and Adaptation: ENVISION case study

Abbas Bradai, Toufik Ahmed and Samir Medjiah
CNRS-LaBRI University of Bordeaux-1
351, Cours de la libération. Talence, 33405
{bradai, tad, medjiah} @labri.fr

*Abstract*—Scalable video coding has drawn great interest in content delivery in many multimedia services thanks to its capability to handle terminal heterogeneity and network conditions variation. In our previous work, and under the umbrella of ENVISION, we have proposed a playout smoothing mechanism to ensure the uniform delivery of the layered stream, by reducing the quality changes that the stream undergoes when adapting to changing network conditions. In this paper we study the resulting video quality, from the final user perception under different network conditions of loss and delays. For that we have adopted the Double Stimulus Impairment Scale (DSIS) method. The results show that the Mean Opinion Score for the smoothed video clips was higher under different network configuration. This confirms the effectiveness of the proposed smoothing mechanism.

## I. INTRODUCTION

The popularity of digital video content together with the increasing number of connected end-users through heterogeneous networks motivates the utilization of Peer-to-Peer (P2P) networks. In these networks, the ability to provide a high level of user experience is a challenging task. This task is supported by effective evaluation of service quality. In order to realize such a quality evaluation service, an assessment model is required to bridge relevant influencing factors with end users' experience.

The typical approaches to video quality assessment are mostly based on Quality of Service (QoS). QoS is referred as a performance measure from the perspective of system performance or reliability. However, QoS metrics do not take into account human users' perception, therefore there is a lack of correlation between QoS and users' opinions. The concept of Quality of Experience (QoE) has been introduced [1] to fill this gap between service performance and user experience. QoE is commonly defined as the overall acceptability of a service as subjectively perceived by end users. Using models based on the concept of QoE, it is possible to measure the extent to which a user-centric service (e.g. video content distribution) achieves its objectives.

In our previous works [2][3][4], under the umbrella of the European project ENVISION [5], we discussed several complexities involved in layered streaming over P2P networks due to bandwidth fluctuation and peer's unreliability. In this case, the correct decision regarding the selection and the scheduling of the layers is crucial: how many layers to request, in which order and from which peer? In [4], we proposed amplitude and frequency reduction (Figure 1) to ensure the smooth quality of the stream and bandwidth allocation schemes for layered streaming systems for P2P networks in [3].

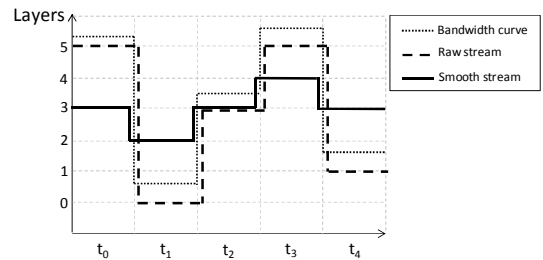

(a) Smoothing by amplitude reduction

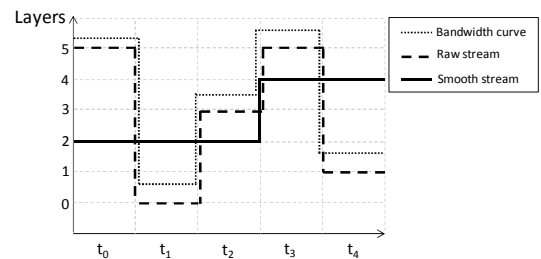

(b) Smoothing by frequency reduction

Figure 1: Variation in quality level (source [4])

The amplitude reduction mechanism aims to reduce the transition from the higher layer to very lower layer (amplitude reduction) and vice versa. The video quality assessment for amplitude reduction can be performed by examining the optimal trade-off between the temporal and SNR scalability. The video basement method such as Double Stimulus Impairment Scale (DSIS) [6] can be used with different frame rate and quantisation parameter but having fixed spatial resolution. The experiment can be performed for different content type such as content having high motion, low motion etc.

Similarly, the subjective evaluation can be performed for frequency reduction mechanism. The goal of frequency reduction is to reduce the number of changes from one layer to another that affects user's perception of stream as described in [4].

## II. STATE OF THE ART

Human perception of different temporal resolutions has been investigated for a relatively long time. Especially, much work has been done to find the minimum acceptable frame rates of video stimuli for various tasks such as target tracking, target detection/recognition, lip reading, orientation judgment, etc. [7]. As for the scenario of video consumption, various factors affect the perceived quality of different frame rates, e.g. content type, viewing condition, display type, and observer characteristics.

For a certain target bit rate, once the spatial resolution is fixed, the temporal resolution can be increased only at the cost of decreased frame quality. The video assessment is performed in order to examine the optimal trade-off between these two dimensions in terms of perceived quality of the final video sequence. Traditionally, it is believed that a high frame rate is more important for content with fast motion than a high frame quality, which is supported by [8]. In [8], subjective experiments were conducted for video sequences encoded by using three different codec's (i.e. the Sorenson codec 2.1, H.263+ and a wavelet-based codec) for eight content types. For a fixed resolution of 352 × 240 pixels, three frame rates (10, 15 and 30 Hz) were considered. Overall, a frame rate of 15 Hz was most preferred across different coding conditions. However, content-dependence was observed, i.e. for content with slow (or fast) motion, preference of a frame rate of 10 Hz (or 30 Hz) was nearly as high as 15 Hz. Similarly, a double stimulus continuous quality scale (DSCQS) experiment in [WSV03] compared H.263+ sequences coded at three different frame rates (7.5, 15 and 30 Hz) with five quantisation parameter (QP) values, for a fixed spatial resolution of 320 × 192 pixels. The results showed that, for slow motion content, subjective quality degradation due to frame rate reduction was only minor.

Different assessment methods have been recommended by the ITU [9], namely (a) The double-stimulus impairment scale (DSIS) [10], (b) The double stimulus continuous quality-scale (DSCQS) [11], (c) Single-stimulus (SS) method [12], (d) Stimulus-comparison method [13], (e) Single stimulus continuous quality evaluation (SSCQE) [14], (f) Simultaneous double stimulus for continuous evaluation method (SDSCE)[15].

In this paper we study the video quality from the final user perception for the SVC content smoothing in different network conditions of loss and delay, in ENVSION content distribution and adaptation framework. In this study we adopt the DSIS method, that will be described in detail in III.B. The rest of the paper is organized as follows: section III describes the evaluation environment, scenarios and methodology. Section VI presents the main obtained results and section V concludes this paper.

## III. EVALUATION SCENARIOS AND METHODOLOGY

### A. Evaluation environment and scenarios

The studied smoothing mechanisms was developed under the ENVISION framework. ENVISION proposes a cross-layer solution where the problem of supporting demanding services is solved cooperatively by service providers, ISPs, users and the applications themselves. To reach this goal, ENVISION relies on CINA [16] interface as a means for cooperation between applications and ISPs for realizing demanding multimedia services.

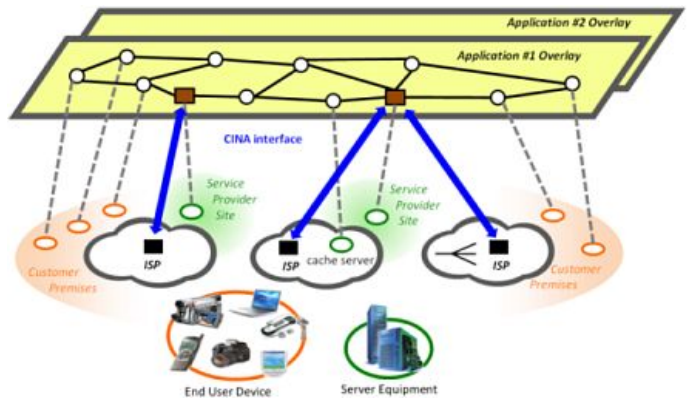

Figure 2: ENVISION overview

In order to permit subjective user evaluation, the streaming application was run under different network and overlay scenarios in ENVISION testbed [17]:

- Use of CINA static costs: The overlay topology construction algorithm may establish overlay connections randomly (*Random policy*), or taking into account the CINA static costs associated with the topology and remaining unchanged in the event of congestion (*RoutingCost*, *RoutingCostLastHop* policies).

- Use of CINA dynamic costs: The overlay content adaptation functions may respond to increased delay or decreased throughput caused in the event of congestion over an overlay link taking into account the CINA dynamic costs (*Latency* and *LatencyLastHop* policies).

- Increasing overlay versus network performance: The overlay topology construction algorithm may establish overlay connections using two different criteria: a) increase the overlay performance by connecting more peers to remote peers with good liveness (*RoutingCost* and *Latency* policies), or b) increase the network performance by reducing the number of connections to remote peers over paths with high CINA costs (*RoutingCostLastHop* and *LatencyLastHop* policies).

*B. Evaluation method*

Evaluation was made using the Double Stimulus Impairment Scale (DSIS) method, also referred as Degradation Category Rating (DCR). In this method videos are shown consequently in pairs. The first one is the reference, and expert is informed about it while the second one is impaired. After their playback, expert is asked to give his opinion on the second one, keeping in mind the first one. During different sessions (each session last up to half an hour) the assessors are shown series of pictures or sequences in random order and impairments covering all the essential combinations. It is important to note that the reference (unimpaired) picture is included in the pictures or sequences to be assessed. Finally, the mean score is calculated at the end of the session for each test condition and test picture.

In our evaluation, the original video is shown for 10 seconds, then the evaluated video for the same 10 seconds. This process was repeated for a total duration of 30 seconds of the video. The scores were expressed over 10, with 0 being the worst score and 10 being the best score.

IV. EVALUATION RESULTS

*A. Performance enhancements at normal network conditions*

At normal network conditions, i.e., no delays and no loss in the network, all the evaluated videos were rated above average for all the metrics used for the overlay construction except those based on *Latency* and *LatencyLastHop*. Figure 3 shows the obtained results. Also, we note for all the evaluated videos, smoothed videos achieved a higher MOS than videos without smoothing.

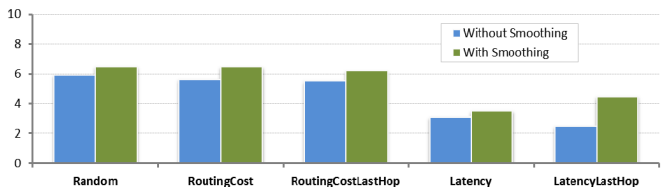

Figure 3: Mean Opinion Score – No delay, no loss

*B. Performance enhancements at the Event of Loss*

In Figure 4, at the event of packet loss (10% loss) in the network, a good streaming experience (video quality) has been achieved for topologies constructed using "Random" and *Latency* metrics (> 6). However, we notice that random topology was more efficient in terms of MOS than topology built using *Latency* metric. Moreover, in this case, smoothed videos have not achieved a higher MOS than videos without smoothing.

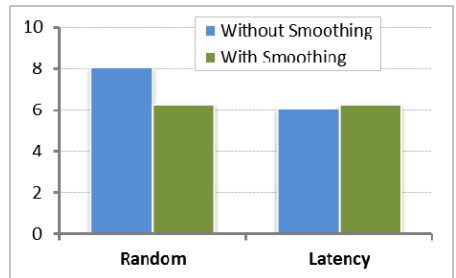

Figure 4: Mean Opinion Score – 10% errors, no delay

*C. Performance enhancements at the Event of Delays*

At the event of delays in the network (delay of 500ms), the resulting streaming experience was rated above average (>5) in the case of different topologies. Figure 5 and Figure 6 show the obtained results. We note that videos from topologies built based on *Latency* and *RoutingCost* achieve the highest MOS (>6). Moreover, in general, smoothed videos are rated slightly higher than videos without smoothing.

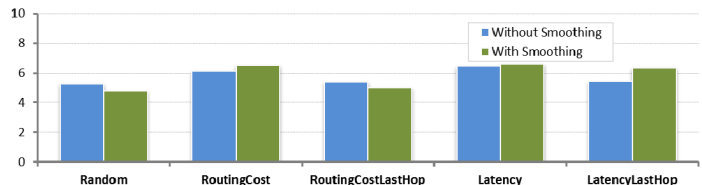

Figure 5: Mean Opinion Score – 500ms delay, no errors

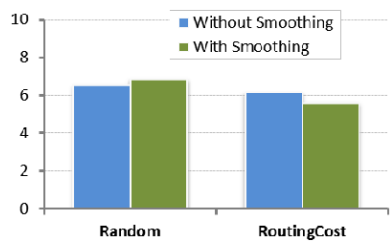

Figure 6: Mean Opinion Score – 1000ms delay, no errors

V. CONCLUSION

In this paper, we have presented a subjective quality measurement study of SVC Live content distribution and adaptation in ENVISION framework. The study was performed using the Double Stimulus Impairment Scale method by LaBRI students with some basic training in content adaptation. Overall, the Mean Opinion Score for the smoothed video clips was higher. In addition, the analysis of some of the low Mean Opinion Scores attributed to some clips has shown that even the slightest of quality fluctuations (a single missed frame) may have a big impact on the value of the Mean Opinion Scores (drop by two points). Thus, the effectiveness of the proposed smoothing mechanism.

**Acknowledgment**

The research leading to these results has received funding from the European Union's Seventh Framework Programme (FP7/2007-2013) in the ENVISION project, grant agreement 248565.

We want also to thank ENVISION partners: ULC, Orange, Alcatel-Lucent, LiveU and Telefonica to their contribution to the design of the FT testbed used in the evaluation scenario.

REFERENCES

[1] Jain R (2004) Quality of experience. IEEE Multimedia 11(1):96–95.
[2] Bradai, A., & Ahmed, T. (2012, June). On the optimal scheduling in pull-based real-time p2p streaming systems: layered and non-layered streaming. In Communications (ICC), 2012 IEEE International Conference on (pp. 1981-1985). IEEE.
[3] Bradai, A., & Ahmed, T. (2012, September). Differenciated bandwidth allocation in P2P layered streaming. In Computer Aided Modeling and Design of Communication Links and Networks (CAMAD), 2012 IEEE 17th International Workshop on (pp. 110-114). IEEE.
[4] Bradai, A., Abbasi, U., Landa, R., & Ahmed, T. (2012). An efficient playout smoothing mechanism for layered streaming in P2P networks. Peer-to-Peer Networking and Applications, 1-17.
[5] http://www.envision-project.org/
[6] Larabi, M. C., Saadane, A., Charrier, C., Fernandez-Maloigne, C., Robert-Inacio, F., & Macaire, L. (2012). Quality Assessment Approaches. Digital Color, 265-306.
[7] Chen, J. Y., & Thropp, J. E. (2007). Review of low frame rate effects on human performance. Systems, Man and Cybernetics, Part A: Systems and Humans, IEEE Transactions on, 37(6), 1063-1076.
[8] Yadavalli, G., Masry, M., & Hemami, S. S. (2003, September). Frame rate preferences in low bit rate video. In Image Processing, 2003. ICIP 2003. Proceedings. 2003 International Conference on (Vol. 1, pp. I-441). IEEE.
[9] International Telecommunication Union: http://www.itu.int/
[10] Larabi, M. C., Saadane, A., Charrier, C., Fernandez-Maloigne, C., Robert-Inacio, F., & Macaire, L. (2012). Quality Assessment Approaches. Digital Color, 265-306.
[11] Inácio, A. P., Cruz, R. S., & Nunes, M. S. (2012). Quality user experience in advanced IP video services. annals of telecommunications-annales des télécommunications, 1-13.
[12] Mantiuk, R. K., Tomaszewska, A., & Mantiuk, R. (2012, December). Comparison of four subjective methods for image quality assessment. In Computer Graphics Forum (Vol. 31, No. 8, pp. 2478-2491). Blackwell Publishing Ltd.
[13] Fiorucci, F., Baruffa, G., & Frescura, F. (2012). Objective and subjective quality assessment between JPEG XR with overlap and JPEG 2000. Journal of Visual Communication and Image Representation, 23(6), 835-844.
[14] Liu, T., Cash, G., Narvekar, N., & Bloom, J. (2012, January). Continuous mobile video subjective quality assessment using gaming steering wheel. In Sixth International Workshop on Video Processing and Quality Metrics for Consumer Electronics (VPQM). Scottsdale, Arizona.
[15] Bosc, E., Le Callet, P., Morin, L., & Pressigout, M. (2013). Visual quality assessment of synthesized views in the context of 3D-TV. In 3D-TV System with Depth-Image-Based Rendering (pp. 439-473). Springer New York.
[16] http://www.envision-project.org/cina/index.html
[17] http://www.envision-project.org/deliverables/envision-d6-2-v2d-public.pdf